\documentclass{egpubl}
\usepackage{eurovis2020}

\SpecialIssuePaper         % uncomment for final version of Computer Graphics Forum, special issue
\usepackage[T1]{fontenc}
\usepackage{dfadobe}  

\usepackage{cite}  % comment out for biblatex with backend=biber

\usepackage{amsfonts}
\usepackage{amsmath}
\usepackage{amssymb}
\usepackage{stmaryrd}

\usepackage{color}
\usepackage{dblfloatfix}    % To enable figures at the bottom of page

\BibtexOrBiblatex
\electronicVersion
\PrintedOrElectronic
\ifpdf \usepackage[pdftex]{graphicx} \pdfcompresslevel=9
\else \usepackage[dvips]{graphicx} \fi

\graphicspath{{Figures/}{./}} % where to search for the images

\usepackage{egweblnk}
\title[HypoML: Visual Analysis for Hypothesis-based Evaluation of Machine Learning Models]%
      {HypoML: Visual Analysis for\\Hypothesis-based Evaluation of Machine Learning Models}

\author[Q. Wang, et al.]
{\parbox{\textwidth}{\centering%
        Qianwen Wang$^{1}$,
        William Alexander$^{2}$,
        Jack Pegg$^{2}$,
        Huamin Qu$^{1}$, and
        Min Chen$^{2}\,$\orcid{0000-0001-5320-5729}
    }
        \\
{\parbox{\textwidth}{\centering%
    $^1$ Hong Kong University of Science and Technology, Hong Kong, China
    \quad and \quad
    $^2$ University of Oxford, UK
    }
}
}
\begin{document}

% uncomment for using teaser
\teaser{
  \centering
  \vspace{-4mm}
  \includegraphics[width=180mm]{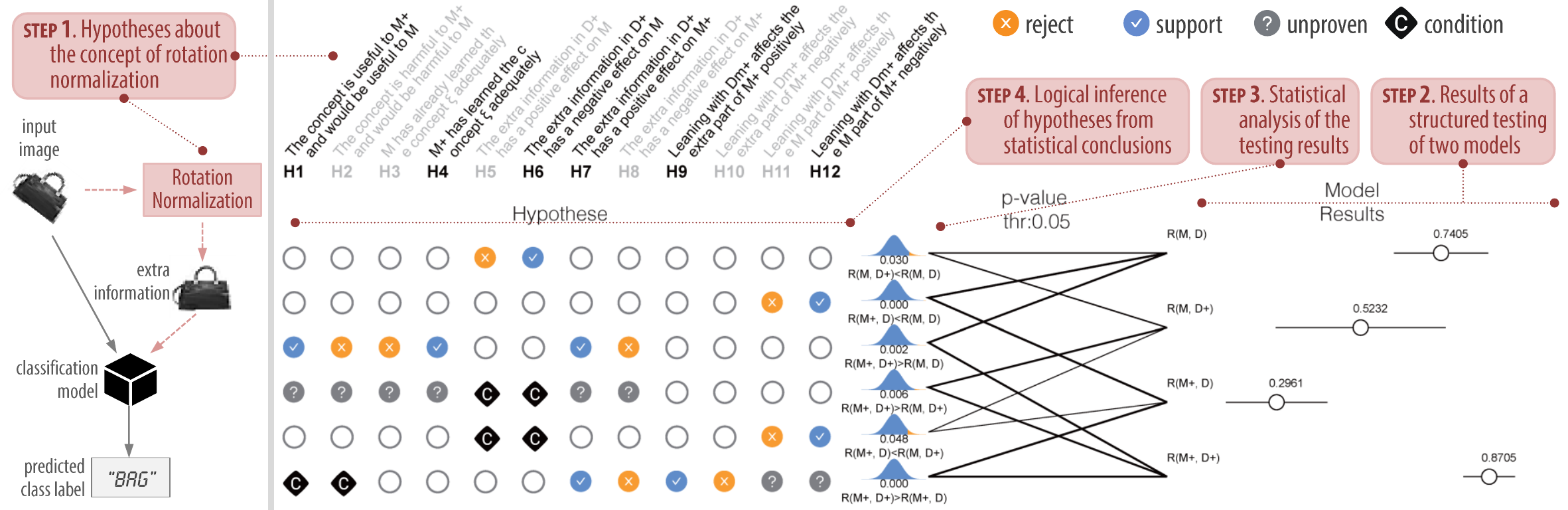}
  \caption{One may wish to know if a classification model is rotation-invariant. If the model is not so, one may use another model that can detect the rotation angle of an object or perform some rotation normalization. The detected rotation angle or normalized image is a piece of extra information about a ``concept'' that the original model may or may not know. With HypoML, one can conduct a set of structured tests, obtained automated statistical and logical analysis of the results, and visualize the conclusions about the hypotheses related to the concept.}
  \label{fig:HypoML}
 }

\maketitle
%-------------------------------------------------------------------------
\begin{abstract}%
In this paper, we present a visual analytics tool for enabling hypothesis-based evaluation of machine learning (ML) models.
We describe a novel ML-testing framework that combines the traditional statistical hypothesis testing (commonly used in empirical research) with logical reasoning about the conclusions of multiple hypotheses.
The framework defines a controlled configuration for testing a number of hypotheses as to whether and how some extra information about a ``concept'' or ``feature'' may benefit or hinder a ML model.  
Because reasoning multiple hypotheses is not always straightforward, we provide HypoML as a visual analysis tool, with which, the multi-thread testing data is transformed to a visual representation for rapid observation of the conclusions and the logical flow between the testing data and hypotheses.
We have applied HypoML to a number of hypothesized concepts, demonstrating the intuitive and explainable nature of the visual analysis.
%    
%-------------------------------------------------------------------------
%  ACM CCS 1998
%  (see https://www.acm.org/publications/computing-classification-system/1998)
% \begin{classification} % according to https://www.acm.org/publications/computing-classification-system/1998
% \CCScat{Computer Graphics}{I.3.3}{Picture/Image Generation}{Line and curve generation}
% \end{classification}
%-------------------------------------------------------------------------
%  ACM CCS 2012
%   (see https://www.acm.org/publications/class-2012)
%The tool at \url{http://dl.acm.org/ccs.cfm} can be used to generate
% CCS codes.
%Example:
% \begin{CCSXML}
% <ccs2012>
% <concept>
% <concept_id>10010147.10010371.10010352.10010381</concept_id>
% <concept_desc>Computing methodologies~Collision detection</concept_desc>
% <concept_significance>300</concept_significance>
% </concept>
% </ccs2012>
% \end{CCSXML}
%
% \ccsdesc[300]{Computing methodologies~Collision detection}
% \ccsdesc[300]{Hardware~Sensors and actuators}
% \ccsdesc[100]{Hardware~PCB design and layout}
%
% \printccsdesc   
\end{abstract}

% ====================
\section{Introduction}
\label{sec:Introduction}
In computer vision, data mining, and machine learning (ML), a \emph{feature} is a measurable variable that characterizes a particular kind of property or attribute of a data object (e.g., an image, a time series, a multivariate record, etc.).
% Many technical solutions in these fields rely on model-developers' knowledge about various features that may potentially be used as the building blocks of an ML model, including feature engineering \cite{Domingos:2012:CACM}, a bag of words \cite{X,Y}, decision tree learning \cite{X}, support vector machine \cite{X}, and so on.
Many technical solutions in these fields heavily rely on model-developers' knowledge about various features and include human-centric feature engineering as a critical process in a model development workflow~\cite{elder1996machine, Alpaydin:2010:IML:1734076}.

% ----------
\begin{figure*}[t]
  \centering
  \includegraphics[width=160mm]{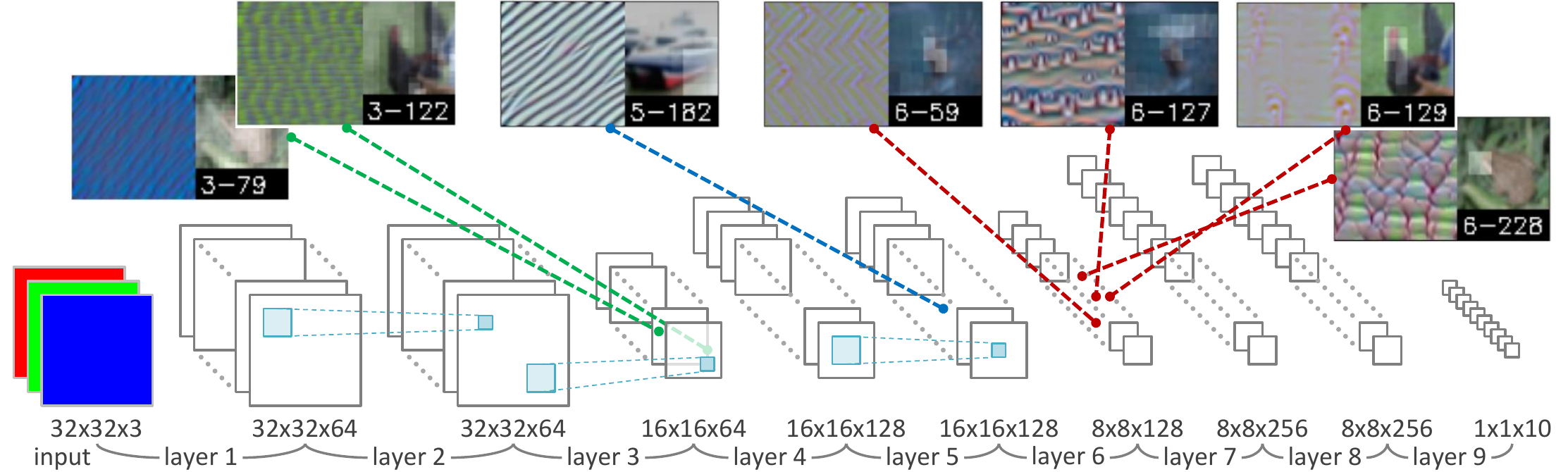}
  \caption{
  Gradient ascent~\cite{springenberg2014striving} can help model-developers observe the pattern that a specified neuron has learned. However, even a small CNN has a huge number of neurons waiting to be inspected, while many patterns shown are not semantically interpretable. Meanwhile, model-developers are often unable to determine whether a pattern is useful or harmful.
%   In deep learning, model-developers often use gradient ascent plots to inspect the signal patterns at different neurons. The figure shows a convolution neural network (CNN) and several of its gradient ascent plots. Some plots indicate the CNN may have learned certain features. Even for this relatively small CNN, there would be many thousands of such plots waiting to be inspected, while many patterns shown are not semantically interpretable. When a gradient ascent plot reveals an interesting pattern, it is not certain if it is useful or harmful.
  }
  \label{fig:GradAct}
  \vspace{-4mm}
\end{figure*}
% ----------

On the other hand, some other technical solutions were designed to minimize the dependence on the human knowledge of potentially useful features.
For example, in deep learning, neural networks are typically expected to learn how to extract a good number of useful features automatically~\cite{lecun2015deep}.
At the same time, there have also been concerns that some so-called ``useful'' features may be actually harmful because they contribute towards undesirable biases \cite{zhang2017achieving, Kilbertus2017avoiding}.
Inevitably, model-developers have been interested in what features may have or have not been learned by an ML model.
A class of visualization techniques, such as neuron activation plot, filter plot, gradient ascent plot~\cite{springenberg2014striving}, Deconvolution~\cite{Zeiler:2014:ECCV}, and their variants, have been widely used by developers of neural networks to observe neurons.
% , typically at the higher layers.
% For a deep neural network, there can still be hundreds and thousands of activation plots at the higher layers.
Since a neural network typically consists of a huge number of neurons, the visual observation may encounter several obstacles, including time demand for viewing all neurons that may reveal some features, subjectivity and memory limitation of an observer, and uncertainty about the semantic meaning of an observed feature.
More importantly, while most model-developers have a non-trivial amount of knowledge about features that are potentially useful or harmful, their initiatives are limited to searching for patterns in many thousands of neuron-based plots and speculating if a feature has been learned.

In this work, we propose a new visual analytics approach that enables model-developers to use their knowledge and initiatives in hypothesising and evaluating if any feature may be useful or harmful, if such a feature is learned by a model, and how it may affect a learned model. In particular, we outline a framework for testing such hypotheses systematically, and describe the underlying statistical and logical analysis for inferring conclusions about multiple hypotheses from multiple sets of testing results.
Because many model-developers may not be familiar with or remember the underlying statistical and logical analysis, we develop a visual analytics tool, HypoML, for carrying out analysis as well as for depicting the flow of inference (Figure \ref{fig:HypoML}), facilitating rapid observation of the conclusions and the logical flow between the testing data and hypotheses.
We have made HypoML available as open-source software, a demo is available at \url{https://hypoml.bitbucket.io/} and the source code is available at \url{https://bitbucket.org/hypoML/hypoml.bitbucket.io}.
% \textcolor{blue}{April: do you have an address?}

The term ``feature'' typically implies a piece of information contained in the original input data.
Since HypoML can also be used to test a hypothesis about a piece of information that may not be part of the original data, we will use the term ``concept-based hypotheses'' to describe what to be tested with HypoML.

% --------------------
\begin{figure*}[t]
  \centering
  \includegraphics[width=180mm]{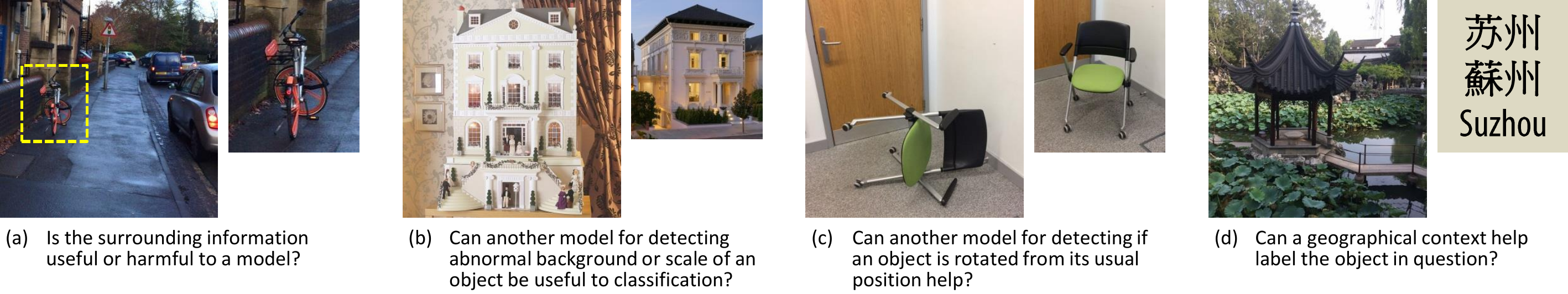}
  \caption{During the development of ML models, model-developers usually have many hypotheses about whether certain extra information (e.g., location, date, time, etc.) or certain preprocessing methods (e.g., cropping, histogram normalization, etc.) would help a model.} 
  \label{fig:Concepts}
\end{figure*}

% ====================
\section{Related Work}
\label{sec:RelatedWork}
While machine learning (ML) has an important role to play in visualization and visual analytics \cite{Endert:2017:CGF}, almost every aspect of ML processes can benefit from visualization as shown by a recently established ontology VIS4ML \cite{Sacha:2019:TVCG}.
In general, when model-developers observe some phenomena in an ML process, such as its training and testing data, results, the inner states of a model, and the provenance of the learning process, they acquire new information to inform their various decisions that affect the ML process. As demonstrated quantitatively by Tam et al. \cite{Tam:2017:TVCG}, a model-developer can contribute a huge amount of knowledge (measured in bits) to an ML process through the use of visualization.
This work focuses on the evaluation stage of ML workflows.

Methods for evaluating ML models can be categorized into two main classes: black-box analysis and white-box analysis.
Here we focus our review of the previous works on model evaluation that feature visualization techniques.
More comprehensive surveys on using visualization for ML can be found in the works of Zhang and Zhu~\cite{Zhang_2018} and Hohman et al.~\cite{Hohman:2020:TVCG}. 

Black-box analysis enables users to investigate and evaluate ML models without knowing the internal working mechanism.
Statistic metrics (e.g., accuracy, recall), ROC curve, and confusion matrices are widely-used black-box analysis and have commonly been provided as built-in functions in machine learning environments.
To aid the aggregated statistical analysis, researchers recently proposed visualization techniques to support black-box evaluation of ML models~\cite{amershi2015modeltracker, krause2016interacting, ren2017squares, zhang2019manifold}.
% % Prospector helps data analysts to understand the prediction of a specific instance through interactively modifying the feature values.
% % ModelTracker~\cite{amershi2015modeltracker} visualizes the prediction scores of each data instance and enables model-developers to conduct an instance-level analysis.
For example, Squares~\cite{ren2017squares} juxtaposes a set of histograms to present an instance-level visualization for models in multi-class classification problems.
Manifold~\cite{zhang2019manifold} employs a scatterplot-based visual technique to assist in the comparison between multiclass classifiers. 
However, these techniques focus mainly on visualizing model performances and offer limited support for model-developers to ask in-depth questions about the model or the experiment results, or to evaluate specific hypotheses in a statistically-meaningful way.

White-box analysis, on the contrary, opens the black box and displays the internal states of ML models.
A number of visualization tools have been proposed to support white-box analysis of different ML models, including MLP~\cite{rauber2017visualizing}, CNNs~\cite{liu2017towards,Kahng2018activis,Pezzotti2018deepeyes,rauber2017visualizing,liu2018deeptracker}, deep generative models~\cite{liu2018DGM,wang2018ganviz, Kahng2019ganlab}, and RNNs~\cite{ming2017understanding,strobelt2018lstmvis}.
Although these tools have utilized some of the most sophisticated visual representations and have assisted model-developers in evaluating, understanding, and explaining their models, comprehending a huge number of high-dimensional internal states is naturally challenging for humans.

In addition, researchers proposed techniques to summarize information about internal variables and present the summary information visually.
Salience-based methods, such as CAM~\cite{zhou2016cam}, Grad-CAM~\cite{Selvaraju2017gradcam}, and guided back propagation~\cite{springenberg2014striving}, identify discriminative regions in the input image and thus highlight important features for a certain prediction.
However, these salience-based methods can only offer explanations for specific predictions but cannot confirm whether or not a concept has been learned.
To offer instance-independent explanation, Yosinski et al. employed gradient ascent plots~\cite{yosinski2015understanding} to depict the patterns that an individual neuron has learned.
\autoref{fig:GradAct} illustrates a small selection of gradient ascent plots being observed in conjunction with a CNN.
However, even for such a simple model, there are a huge number of neurons, it is impossible for model-developers to conduct a full examination.
Moreover, the depicted pattern would largely be a hunch, but not a proof that a certain concept is useful or not to the classification task.
Perhaps the most relevant to our work is TCVA~\cite{been2017TCAV},  which learns human-friendly concepts from an already trained model and conducts hypothesis testing. 
However, TCVA requires a time-consuming process to label the concept across the whole dataset.

In this work, we propose a novel ML-testing framework that combines black-box and white-box analysis.
Whether an ML model has learned a concept or feature is a typical ``internal problem'' that is to be investigated using white-box analysis.
The new framework allows model-developers to investigate ``internal problems'' in a manner of black-box analysis.

% ====================
\section{Concept-Based Testing of ML Models}
\label{sec:Testing}
Let $\textrm{M}$ be a machine learned (ML) model that transforms an input data object $d \in \mathrm{D}$ to an output decision that may be of a classification label, or a prediction.
A \emph{concept} $\xi$ is a variable that is not explicitly defined in $d_i$, but is hypothesized by an ML model-developer that $\xi$ would be useful or harmful to the quality of the output decision should $\textrm{M}$ be able to access some extra information about $\xi$.
Figure \ref{fig:Concepts} shows several examples of concepts.
We can observe that some concepts may be extracted from the original data objects using known techniques, while it may be almost impossible to infer some other concepts from the data objects.

As long as $\textrm{M}$ has a finite number of constructs (e.g., neurons or tree nodes) or receives input data with finite informative dimensions, there will always be some concepts that $\textrm{M}$ cannot learn.
Inevitably, most model-developers will have questions about some concepts in relation to a learned model $\textrm{M}$.
For example, considering the examples in Figure \ref{fig:Concepts}, one may ask:
\begin{enumerate}
    \item[a.] Would having an extended field of view be useful for recognizing an object captured from a less ideal viewing angle?
    \item[b.] Would another model for detecting an anomalous background or some scale inconsistency be useful to differentiate a toy from a real building?
    \item[c.] Would another model that is able to detect an object in an unusual position and estimate the rotation angle be useful to the recognition of the object?
    \item[d.] Would having additional information about a geographical context  improve the accuracy of building recognition?
\end{enumerate}

One can easily imagine many other questions about different types of extra information, such as different meta-data, multiple data capture modalities, and various pre-processing techniques.
All these questions are essentially hypotheses.
Just as in psychology, healthcare, social science, and many other disciplines, one can conduct experiments to evaluate such hypotheses.
Indeed, one can test ML models against many thousands of data objects in comparison with tens of stimuli in typical empirical studies.

Because model testing is a routine operation in ML, it is desirable to establish a structured method such that many model-developers can adopt the same method and produce comparable testing results.
Because the above definition of \emph{concept} is relatively broad, developers of different ML models in various applications can benefit from open source software or commercial systems for supporting such a structured testing method.  

% --------------------
\begin{figure*}[t]
  \centering
  \includegraphics[width=160mm]{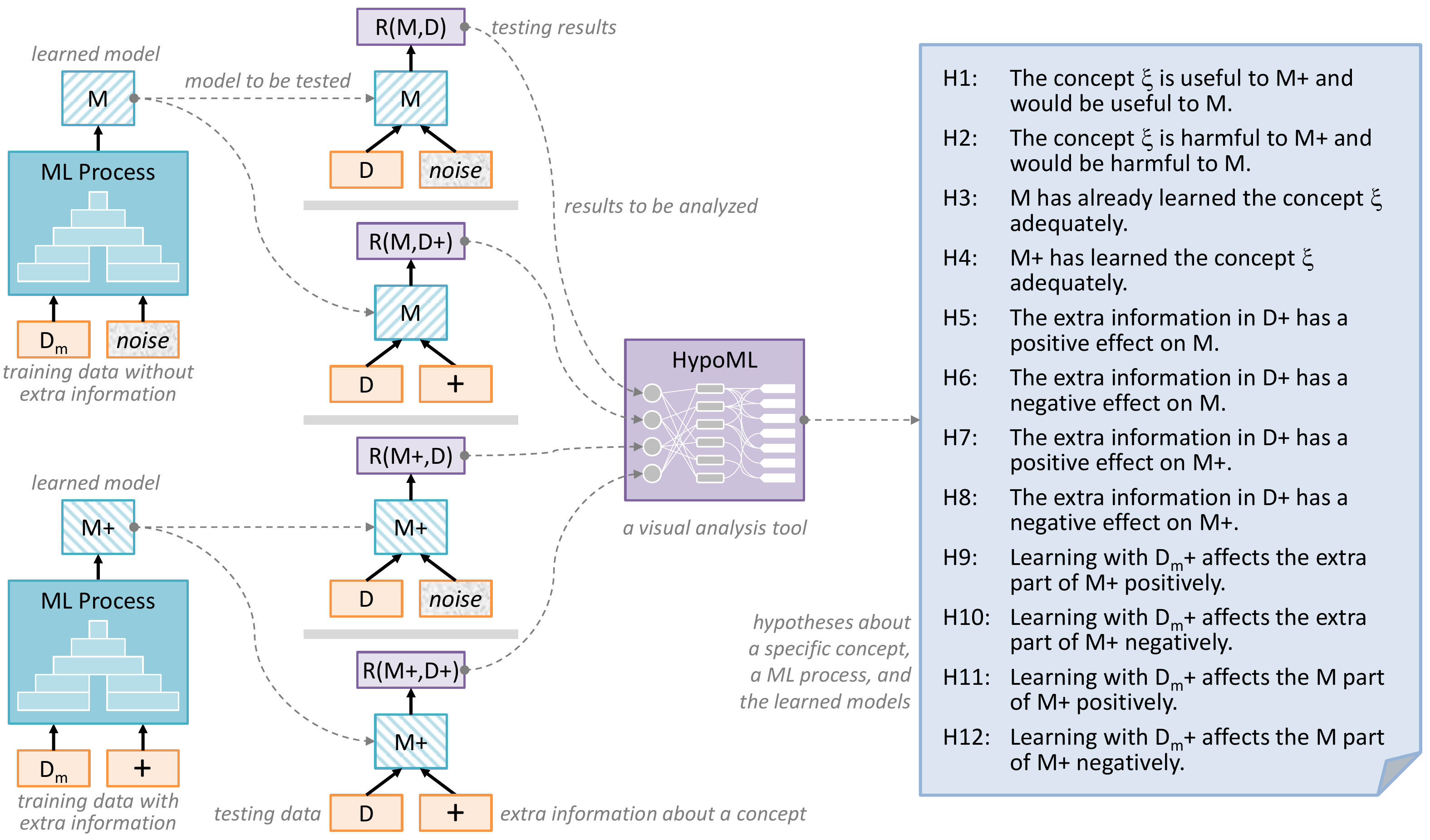}
  \caption{An illustration of the structured testing method proposed in this work. The concept to be tested is encoded as extra information to accompany the original data. Two models, $\mathrm{M+}$ and $\mathrm{M}$, are trained with and without extra information. Both models are then tested using two different types of testing data, one with extra information and one without. The four sets of results are then analyzed by the HypoML tool against 12 hypotheses. HypoML presents the analytical conclusions using visualization as shown in Figure \ref{fig:HypoML}.}
  \label{fig:Testbed}
  \vspace{-4mm}
\end{figure*}

Figure \ref{fig:Testbed} illustrates the framework for concept-based hypothesis testing.
Given an ML process and a training and testing dataset, a model-developer is interested to know how some extra information about a concept may affect the ML process and the learned model.
The framework thus requires the developer to invoke two ML processes that receive two pieces of input data.
As shown on the left of Figure \ref{fig:Testbed}, both processes take the original training data $\mathrm{D_m}$ as one piece of the data.
For the other piece, one process takes random noise as its input, while the other takes extra information about a concept (denoted by the sign ``$+$'').

Following the same procedure for model training, the two ML processes generate two learned models, $\mathrm{M}$ and $\mathrm{M+}$, respectively. The framework then requires the model-developer to test each model with two runs.
As illustrated in the middle column of Figure \ref{fig:Testbed}, one testing run uses testing data $\mathrm{D}$ that does not have extra information, while the other run uses testing data $\mathrm{D+}$ that include extra information.
The two runs with $\mathrm{M}$ thus produce two sets of results, $\mathrm{R_{M,D}}$ and $\mathrm{R_{M,D+}}$, while the two runs with $\mathrm{M+}$ produce $\mathrm{R_{M+,D}}$ and $\mathrm{R_{M+,D+}}$.
Because evaluating an ML model typically involves testing many thousands of data objects, some computational analysis of the four sets of results will be necessary.   

HypoML is designed to support the computational analysis.
In particular, it provides statistical and logical analysis for evaluating a set of hypotheses.
The statistical analysis is based on the well-established method for hypothesis testing, while the logical analysis is formulated in this work for reasoning about the intertwining relationships between 12 hypotheses and 6 statistical conclusions drawn from different pairs of results.
To assist users in understanding such complex relationships, HypoML provides a purposely-designed visual representation, which enables users to trace the conclusion of each hypothesis to related statistical analysis and to the corresponding testing results.

The 12 hypotheses are listed on the right of Figure \ref{fig:Testbed}.
The first two hypotheses, $\mathbf{H}_1$ and $\mathbf{H}_2$, are about whether the concept concerned is useful (or harmful) to $\mathrm{M}+$, and would be useful (or harmful) to $\mathrm{M}$.
Although the conclusions for these two hypotheses cannot in principle be both positive, each can also be inconclusive.
We thus follow the convention of hypothesis testing by listing them as separate hypotheses, each can be independently confirmed, rejected, or unproven (inconclusive).

$\mathbf{H}_3$ hypothesizes that model $\mathrm{M}$ has already learned the concept adequately, while $\mathbf{H}_4$ hypothesizes that model $\mathrm{M}+$ has learned the concept adequately.
For $\mathbf{H}_3$, the adverb ``adequately'' implies that the concept can be learned by a model, such as $\mathrm{M}$, without the need for any extra information about the concept.
For $\mathbf{H}_4$, the adverb ``adequately'' implies that $\mathrm{M}+$ would perform worse without the extra information of the concept.

In general, model $\mathrm{M}$ has not been trained with extra information.
It is thus not expected to be affected by any extra information during testing.
However, as a scientific exercise, one cannot take this assumption for granted since one cannot assume that a model template (e.g., an untrained neural network) has always been configured correctly or a training method has always been implemented correctly.
$\mathbf{H}_5$ and $\mathbf{H}_6$ are thus designed to examine whether $\mathrm{M}$ is affected positively or negatively by the extra information during testing.
Because there exists an inconclusive state, they are kept as two separate hypotheses, in a way similar to $\mathbf{H}_1$ and $\mathbf{H}_2$.

When model $\mathrm{M}+$ is trained with extra information, the model may learn new capability from the extra information, while losing some capability that would be learned without the extra information. The vice versa could also be true.
$\mathbf{H}_9$, $\mathbf{H}_{10}$, $\mathbf{H}_{11}$, and $\mathbf{H}_{12}$ are for investigating the trade-off between different parts of $\mathrm{M}+$ in the development of its intelligence.  
Depending on the design of the model template or architecture, the parts of $\mathrm{M}+$ for handling the extra information ($+$ part) and the original information ($\mathrm{D}$) can be quite separated as well as rather integrated.
When the two parts are more integrated, one should consider two parts as functional units rather than geometric or topological regions.
Similarly, we separate $\mathbf{H}_9$ from $\mathbf{H}_{10}$, and separate $\mathbf{H}_{11}$ from $\mathbf{H}_{12}$ because of the inconclusive state in each case.
We also anticipate that more testing and analysis methods may be developed in the future, which may support or reject those apparently-paired hypotheses asymmetrically.
Having separate hypotheses will not hinder such advancement.

% ====================
\section{Statistical and Logical Reasoning of Hypotheses}
\label{sec:Reasoning}
As shown in Figure \ref{fig:Testbed}, HypoML receives four sets of results, namely $\mathrm{R_{M,D}}$, $\mathrm{R_{M,D+}}$, $\mathrm{R_{M+,D}}$, and $\mathrm{R_{M+,D+}}$.
Each set of results is a list of tuples, each of which consists of:
\begin{itemize}
  \item \textsf{id} --- the unique identifier of a data object. The data object may be an image, a feature vector, a multivariate data record, or a more complex data record.
  \item \textsf{ground truth} --- a ground truth label, which can be a nominal value, an integer, a real number, a range, or a data record of a more complex data type (e.g., a time series). 
  \item \textsf{ML label} --- a label generated by an ML model. The label must be of the same data type as \textsf{ground truth}.
  \item \textsf{ML uncertainty} --- an optional value indicating the uncertainty estimated by an ML model equipped with a self-assessment capacity. It is a real number in the range [0, 1] with 1 being the most uncertain. Many ML models may not have any self-assessment capacity, and in such a case, this entry takes the default value 0. Some ML models may return a confidence value, which can easily be converted to uncertainty.
  \item \textsf{correctness} --- This is a value in the range of [0, 1] with 1 indicating absolutely correct, and 0 indicating absolutely incorrect. The value is mostly computed based on \textsf{ground truth} and \textsf{ML label} using a user-defined function. The simplest function can be true (1) if \textsf{ground truth} equals \textsf{ML label}, or 0 otherwise. A more complicated function may feature a distance or similarity metric.  
  \item \textsf{correctness with uncertainty} --- This is used by the statistical analysis and is defined as \textsf{ML uncertainty} $\times$ \textsf{correctness}.
\end{itemize}

Given two sets of results, $\mathrm{R_a}$ and $\mathrm{R_b}$, we assume that the tuples in the two lists are paired, i.e., the \textsf{id} entries are in the same order exactly. We can compare $\mathrm{R_a}$ and $\mathrm{R_b}$ with their accuracy, i.e., the average of \textsf{correctness with uncertainty}.
As testing in ML often shows small variations of accuracy, it is necessary to measure the statistical significance.
HypoML uses paired, two-tail $t$-test for this purpose.
Let us introduce the following notation to denote the possible outcomes of the statistical analysis.

\begin{itemize}
    \item $\mathrm{R_a} \lnapprox \mathrm{R_b}$ --- It is statistically significant that $\mathrm{R_a}$ is lower than $\mathrm{R_b}$.
    \item $\mathrm{R_a} \gnapprox \mathrm{R_b}$ --- It is statistically significant that $\mathrm{R_a}$ is higher than $\mathrm{R_b}$.
    \item $\mathrm{R_a} \approx \mathrm{R_b}$ --- It is statistically insignificant that $\mathrm{R_a}$ is higher or lower than $\mathrm{R_b}$.
    \item $\mathrm{R_a} \lessapprox \mathrm{R_b}$ --- $\mathrm{R_a} \lnapprox \mathrm{R_b}$ or $\mathrm{R_a} \approx \mathrm{R_b}$, but not $\mathrm{R_a} \gnapprox \mathrm{R_b}$.
    \item $\mathrm{R_a} \gtrapprox \mathrm{R_b}$ --- $\mathrm{R_a} \gnapprox \mathrm{R_b}$ or $\mathrm{R_a} \approx \mathrm{R_b}$, but not $\mathrm{R_a} \lnapprox \mathrm{R_b}$.  
\end{itemize}

With four sets of results, there are six pairs of statistical comparison, which are labelled as $A_1, A_2, \ldots, A_6$. Each analytical conclusion $A_i$ may support or reject some of the 12 hypotheses $\mathbf{H}_1, \mathbf{H}_2, \ldots, \mathbf{H}_{12}$, but not all.
For example the analysis $A_1$, which compares $\mathrm{R_{M+,D+}}$ and $\mathrm{R_{M,D}}$, can inform the evaluation of $\mathbf{H}_1$ and $\mathbf{H}_2$.
If $\mathrm{R_{M+,D+}}$ is statistically better than $\mathrm{R_{M,D}}$, i.e., $\mathrm{R_{M+,D+}} \gnapprox \mathrm{R_{M,D}}$, we can draw a conclusion that $A_1$ supports $\mathbf{H}_1$ and rejects $\mathbf{H}_2$.
If $\mathrm{R_{M+,D+}} \lnapprox \mathrm{R_{M,D}}$, $A_1$ supports $\mathbf{H}_2$ and rejects $\mathbf{H}_1$.
If $\mathrm{R_{M+,D+}} \approx \mathrm{R_{M,D}}$, $A_1$ returns an unproven (inconclusive) verdict about $\mathbf{H}_1$ and $\mathbf{H}_2$.

With some careful reasoning, we can observe that $A_1$ can also inform the evaluation of $\mathbf{H}_3$, $\mathbf{H}_4$, $\mathbf{H}_7$, and $\mathbf{H}_8$.
While $A_2$ can inform the evaluation of $\mathbf{H}_1$, $\mathbf{H}_2$, $\mathbf{H}_3$, $\mathbf{H}_4$, $\mathbf{H}_7$, and $\mathbf{H}_8$, but it can only do so subject to that some other hypotheses have already been confirmed or rejected.
Table \ref{tab:Relations} summaries the relations between the six sets of statistical analysis $A_1, A_2, \ldots, A_6$ and the 12 hypotheses $\mathbf{H}_1, \mathbf{H}_2, \ldots, \mathbf{H}_{12}$.

\begin{table}[t]
  \centering
  \begin{tabular}{@{}l|@{\hspace{1mm}}c@{\hspace{1mm}}|l@{}}
  \textbf{Analysis} & \textbf{Condition} & \textbf{Hypothesis}\\
  \hline
  $A_1$: $\mathrm{R_{M+,D+}}$ v. $\mathrm{R_{M,D}}$
    & & $\mathbf{H}_1$, $\mathbf{H}_2$, $\mathbf{H}_3$, $\mathbf{H}_4$, $\mathbf{H}_7$, $\mathbf{H}_8$\\
  $A_2$: $\mathrm{R_{M+,D+}}$ v. $\mathrm{R_{M,D+}}$ 
    & $\mathbf{H}_5$, $\mathbf{H}_6$
    & $\mathbf{H}_1$, $\mathbf{H}_2$, $\mathbf{H}_3$, $\mathbf{H}_4$, $\mathbf{H}_7$, $\mathbf{H}_8$\\
  $A_3$: $\mathrm{R_{M+,D+}}$ v. $\mathrm{R_{M+,D}}$
    & $\mathbf{H}_1$, $\mathbf{H}_2$
    & $\mathbf{H}_7$, $\mathbf{H}_8$, $\mathbf{H}_9$, $\mathbf{H}_{10}$, $\mathbf{H}_{11}$, $\mathbf{H}_{12}$\\
  $A_4$: $\mathrm{R_{M+,D}}$ v. $\mathrm{R_{M,D}}$
    & & $\mathbf{H}_{11}$, $\mathbf{H}_{12}$\\
  $A_5$: $\mathrm{R_{M+,D}}$ v. $\mathrm{R_{M,D+}}$
    & $\mathbf{H}_5$, $\mathbf{H}_6$
    & $\mathbf{H}_{11}$, $\mathbf{H}_{12}$\\
  $A_6$: $\mathrm{R_{M,D+}}$ v. $\mathrm{R_{M,D}}$
    & & $\mathbf{H}_5$, $\mathbf{H}_6$\\
  \hline
  \end{tabular}
  \caption{The relations between statistical analysis and hypotheses.}
  \label{tab:Relations}
  \vspace{-4mm}
\end{table}

Clearly, reasoning about these relations is time consuming and error prone.
In order to support the frequent analytical tasks of the developers in testing their ML models, HypoML provides automated logical analysis as well as statistical analysis.
To help describe the logical analysis, we employ some additional notations.
They are:
\begin{itemize}
  \item $\top(S)$ --- The statement $S$ is true.
  \item $\bot(S)$ --- The statement $S$ is false.
  \item $\divideontimes(S)$ --- The statement $S$ is unproven.
  \item $\wedge$ --- Logical conjunction.
  \item $\vee$ --- Logical (inclusive) disjunction.
\end{itemize}

W can now specify the logical inference from $A_1$ as:

\noindent $A_1$: $\mathrm{R_{M+,D+}}$ v. $\mathrm{R_{M,D}}$ may conclude:
\begin{itemize}
    \item $\mathrm{R_{M+,D+}} \gnapprox \mathrm{R_{M,D}}$ $\Longrightarrow$
    $\top(\mathbf{H}_1) \wedge \top(\mathbf{H}_4) \wedge \top(\mathbf{H}_7) \wedge \bot(\mathbf{H}_2) \wedge \bot(\mathbf{H}_3) \wedge \bot(\mathbf{H}_8)$. This reads as $\mathbf{H}_1$, $\mathbf{H}_4$, and $\mathbf{H}_7$ are all true, and $\mathbf{H}_2$, $\mathbf{H}_3$, and $\mathbf{H}_8$ are all false.
    \item $\mathrm{R_{M+,D+}} \lnapprox \mathrm{R_{M,D}}$ $\Longrightarrow$
    $\top(\mathbf{H}_2) \wedge \top(\mathbf{H}_4) \wedge \bot(\mathbf{H}_1) \wedge \bot(\mathbf{H}_3)$.
\end{itemize}

Analysis $A_2$ cannot draw conclusions about $\mathbf{H}_5$ and $\mathbf{H}_6$, but its conclusion may depend on them.
In general, there is a common-sense assumption that neither $\mathbf{H}_5$ nor $\mathbf{H}_6$ is likely to be true.

\noindent $A_2$: $\mathrm{R_{M+,D+}}$ vs. $\mathrm{R_{M,D+}}$ may conclude:
\begin{itemize}
    \item $\mathrm{R_{M+,D+}} \gnapprox \mathrm{R_{M,D+}}$ $\Longrightarrow$\\
    (i) \textbf{if} $\bot(\mathbf{H}_6)$ \textbf{then} $\top(\mathbf{H}_1) \wedge \top(\mathbf{H}_4) \wedge \top(\mathbf{H}_7) \wedge \bot(\mathbf{H}_2) \wedge \bot(\mathbf{H}_3) \wedge \bot(\mathbf{H}_8)$; \textbf{or}\\
    (ii) \textbf{if} $\divideontimes(\mathbf{H}_6)$ \textbf{then} $\top(\mathbf{H}_1) \wedge \top(\mathbf{H}_4) \wedge \top(\mathbf{H}_7) \wedge \bot(\mathbf{H}_2) \wedge \bot(\mathbf{H}_3) \wedge \bot(\mathbf{H}_8)$; \textbf{or}\\
    (iii) \textbf{if} $\top(\mathbf{H}_6)$. This offers an explanation but it is against a common-sense assumption that $\mathbf{H}_6$ is unlikely to be true, and should be treated cautiously.
    \item $\mathrm{R_{M+,D+}} \lnapprox \mathrm{R_{M,D+}}$ $\Longrightarrow$\\
    (i) \textbf{if} $\bot(\mathbf{H}_5)$ \textbf{then} $\top(\mathbf{H}_2) \wedge \top(\mathbf{H}_4) \wedge \top(\mathbf{H}_8) \wedge \bot(\mathbf{H}_1) \wedge \bot(\mathbf{H}_3) \wedge \bot(\mathbf{H}_7)$; \textbf{or}\\
    (ii) \textbf{if} $\divideontimes(\mathbf{H}_5)$ \textbf{then} $\top(\mathbf{H}_2) \wedge \top(\mathbf{H}_4) \wedge \top(\mathbf{H}_8) \wedge \bot(\mathbf{H}_1) \wedge \bot(\mathbf{H}_3) \wedge \bot(\mathbf{H}_7)$; \textbf{or}\\
    (iii) \textbf{if} $\top(\mathbf{H}_5)$. This offers an explanation but it is against a common-sense assumption that $\mathbf{H}_5$ is unlikely to be true, and should be treated cautiously.
\end{itemize}

Because analysis $A_3$ does not compare $\mathrm{M+}$ with $\mathrm{M}$, the conclusion is limited to the context of $\mathrm{M+}$. Mathematically, it is possible for $A_3$ to conclude that the concept is useful in the context of $\mathrm{M+}$, while $A_1$ or $A_2$ concludes that the concept is harmful or is neither useful nor harmful.
Considering this limitation, it is unsafe for this analysis to draw a conclusion about $\mathbf{H}_1$ and $\mathbf{H}_2$. Meanwhile the analysis depends on the conclusions of $\mathbf{H}_1$ and $\mathbf{H}_2$ in a small way.

\noindent $A_3$: $\mathrm{R_{M+,D+}}$ v. $\mathrm{R_{M+,D}}$ may conclude:
\begin{itemize}
    \item $\mathrm{R_{M+,D+}} \gnapprox \mathrm{R_{M+,D}}$ $\Longrightarrow$\\
    (i) \textbf{if} $\top(\mathbf{H}_1)$, \textbf{then} $\top(\mathbf{H}_7) \wedge \top(\mathbf{H}_9) \wedge \bot(\mathbf{H}_8) \wedge \bot(\mathbf{H}_{10})$; \textbf{or}\\
    (ii) \textbf{if} $\divideontimes(\mathbf{H}_1)$, \textbf{then} $\top(\mathbf{H}_9) \wedge \bot(\mathbf{H}_{10})$; \textbf{or}\\
    (iii) \textbf{if} $\bot(\mathbf{H}_1)$, \textbf{then} $\top(\mathbf{H}_{12}) \wedge \bot(\mathbf{H}_{11})$.
    \item $\mathrm{R_{M+,D+}} \lnapprox \mathrm{R_{M+,D}}$ $\Longrightarrow$\\
    (i) \textbf{if} $\top(\mathbf{H}_2)$, \textbf{then} $\top(\mathbf{H}_8) \wedge \top(\mathbf{H}_{10}) \wedge \bot(\mathbf{H}_7) \wedge \bot(\mathbf{H}_9)$; \textbf{or}\\
    (ii) \textbf{if} $\divideontimes(\mathbf{H}_2)$, \textbf{then} $\top(\mathbf{H}_{10}) \wedge \bot(\mathbf{H}_9)$; \textbf{or}\\
    (iii) \textbf{if} $\bot(\mathbf{H}_2)$, \textbf{then} $\top(\mathbf{H}_{10}) \wedge \bot(\mathbf{H}_{9})$. This conclusion is against a common-sense assumption that a useful concept normally should not affect the extra part of M+ negatively, and should be treated cautiously. 
\end{itemize}

Analysis $A_4$ is relatively easy to reason, and it is useful for investigating if the part of model $\mathrm{M}+$ for handling the original data $\mathrm{D}$ becomes less capable due to the training with extra information.

\noindent $A_4$: $\mathrm{R_{M+,D}}$ v. $\mathrm{R_{M,D}}$ may conclude:
\begin{itemize}
    \item $\mathrm{R_{M+,D}} \gnapprox \mathrm{R_{M,D}}$ $\Longrightarrow$
    $\top(\mathbf{H}_{11}) \wedge \bot(\mathbf{H}_{12})$.
    \item $\mathrm{R_{M+,D}} \lnapprox \mathrm{R_{M,D}}$ $\Longrightarrow$
    $\top(\mathbf{H}_{12}) \wedge \bot(\mathbf{H}_{11})$.
\end{itemize}

$A_5$ cannot draw conclusions about $\mathbf{H}_5$ and $\mathbf{H}_6$, but its conclusion may depend on them.
In general, there is a common-sense assumption that neither $\mathbf{H}_5$ nor $\mathbf{H}_6$ is true.

\noindent $A_5$: $\mathrm{R_{M+,D}}$ v. $\mathrm{R_{M,D+}}$ may conclude:
\begin{itemize}
    \item $\mathrm{R_{M+,D}} \gnapprox \mathrm{R_{M,D+}}$ $\Longrightarrow$\\
    (i) \textbf{if} $\bot(\mathbf{H}_6)$ \textbf{then} $\top(\mathbf{H}_{11}) \wedge \bot(\mathbf{H}_{12})$; \textbf{or}\\
    (ii) \textbf{if} $\divideontimes(\mathbf{H}_6)$ \textbf{then} $\top(\mathbf{H}_{11}) \wedge \bot(\mathbf{H}_{12})$; \textbf{or}\\
    (iii) \textbf{if} $\top(\mathbf{H}_6)$. This offers an explanation but it is against a common-sense assumption that $\mathbf{H}_6$ is unlikely to be true, and should be treated cautiously.
    \item $\mathrm{R_{M+,D}} \lnapprox \mathrm{R_{M,D+}}$ $\Longrightarrow$\\
    (i) \textbf{if} $\bot(\mathbf{H}_5)$ \textbf{then} $\top(\mathbf{H}_{12}) \wedge \bot(\mathbf{H}_{11})$; \textbf{or}\\
    (ii) \textbf{if} $\divideontimes(\mathbf{H}_5)$ \textbf{then} $\top(\mathbf{H}_{12}) \wedge \bot(\mathbf{H}_{11})$; \textbf{or}\\
    (iii) \textbf{if} $\top(\mathbf{H}_5)$. This offers an explanation but it is against a common-sense assumption that $\mathbf{H}_6$ is unlikely to be true, and should be treated cautiously.
\end{itemize}

Analysis $A_6$ is the only comparison that may inform the evaluation of $\mathbf{H}_5$ nor $\mathbf{H}_6$. In general, there is a common-sense assumption that neither $\mathbf{H}_5$ nor $\mathbf{H}_6$ is true if the model template or architecture was correctly defined, the correct ML method was followed, and the correct ML process was executed.
When $\mathbf{H}_5$ or $\mathbf{H}_6$ is confirmed, it usually suggests some imperfection of the model template or learning process.
Therefore the conclusions of $A_6$ should not be interpreted as their face values.
However, the evaluation of $\mathbf{H}_5$ nor $\mathbf{H}_6$ is necessary since $A_2$ and $A_5$ depend on them.

\noindent $A_6$: $\mathrm{R_{M,D+}}$ vs. $\mathrm{R_{M,D}}$ may conclude:
\begin{itemize}
    \item $\mathrm{R_{M,D+}} \gnapprox \mathrm{R_{M,D}}$ $\Longrightarrow$
    $\top(\mathbf{H}_5) \wedge \bot(\mathbf{H}_6)$;
    \item $\mathrm{R_{M,D+}} \lnapprox \mathrm{R_{M,D}}$ $\Longrightarrow$
    $\top(\mathbf{H}_6) \wedge \bot(\mathbf{H}_5)$.
\end{itemize}

Because the dependency among the six sets of analysis, the computation of the logical inference must follow an appropriate order, which is summarized as follows:

\noindent STEP 0: Initialise the indicator of each hypothesis to 0.

\noindent STEP 1: Compute the six comparative values, i.e., $A_1, A_2, \ldots, A_6$, in terms of $\lnapprox$, $\gnapprox$, and $\approx$, based on statistical analysis.

\noindent STEP 2: Compute the logical inference (i.e., in terms of $\top, \bot, \divideontimes$) based on $A_1$, $A_4$, $A_6$. For each true statement, i.e., $\top(\mathbf{H}_i)$, add $+1$ to the indicator of $\mathbf{H}_i$. For each false statement, i.e., $\bot(S)$, add $-1$ to the indicator of $\mathbf{H}_i$.

\noindent STEP 3: Compute the indicators based on $A_2$, $A_5$.

\noindent STEP 4: Compute the indicators based on $A_3$.

\noindent STEP 5: Then display each indicator based on positive or negative values. HypoML displays each hypothesis according to its indicator in three states: $>\!\!0$ (confirmed), 0 (unproven), $<\!\!0$ (rejected).

% ====================
\section{Visual Analysis of Hypotheses}
\label{sec:VisualAnalysis}

\begin{figure}[t]
    \centering
    \includegraphics[width=\linewidth]{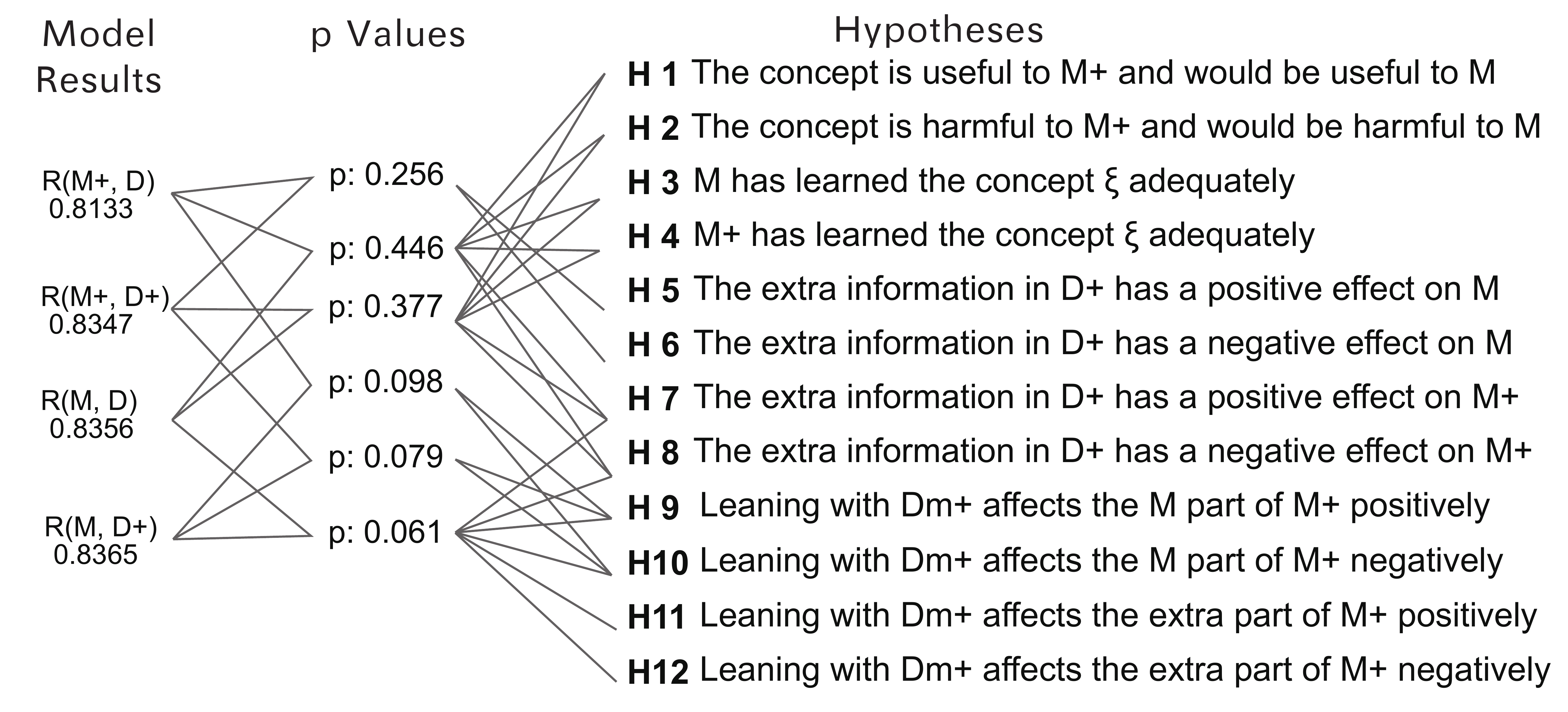}
    \caption{The analytical workflow from testing results to statistical analysis and then logical inference of hypothesis. As a basic visual design, it has a number of shortcomings.}
    \label{fig:workflow}
    \vspace{-4mm}
\end{figure}

\begin{figure}[t]
    \centering
    \includegraphics[width=\linewidth]{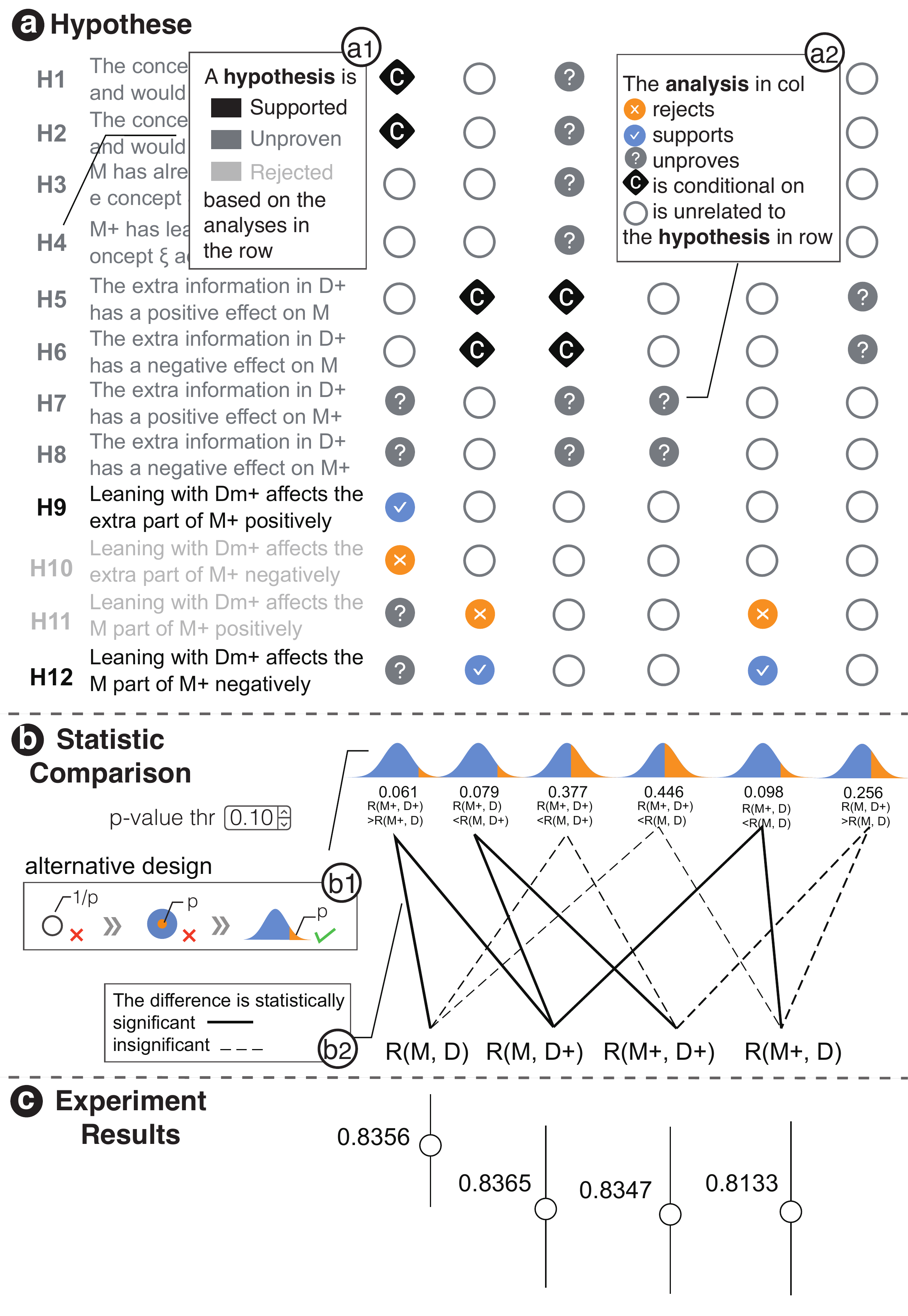}
    \caption{The vertical version of the HypoML interface. In comparison with the basic design in \autoref{fig:workflow}, it is much easier for a user to have an overview of the analytical flow, while acquiring quickly the conclusions of different hypotheses.
    A horizontal version, which is more suitable for wide-screen monitors, is shown in \autoref{fig:HypoML}.}
    \label{fig:interface_vertical}
    \vspace{-4mm}
\end{figure}

\autoref{fig:workflow} shows a typical workflow of the proposed hypotheses testing. 
To start with, model-developers conduct experiments and obtain four sets of results, i.e., $\mathrm{R_{M,D}}$, $\mathrm{R_{M,D+}}$, $\mathrm{R_{M+,D}}$, and $\mathrm{R_{M+,D+}}$.
HypoML then performs six sets of statistical analysis by comparing each pair of the results.
Based on the statistical analysis, HypoML makes logical inference about the twelve hypotheses, deciding whether a hypothesis should be supported or rejected.

It is helpful for model-developers to make quick observation about the analysis and conclusions.
It will also be useful for the model-developers to convey the outcomes of the test to other stakeholders, such as users of the ML models being evaluated.
It can be difficult for some model-developers and many of ML users to remember and reason the complicated relationships among experiment results, statistic and logical analysis, and multiple hypotheses. 
Therefore, an effective visual representation is necessary.
The bipartite graph shown in \autoref{fig:workflow} is a straightforward solution but it exhibits several shortcomings that hinder efficient information acquisition and effective information dissemination.

One main shortcoming is the cluttered links between the six statistical comparisons and the twelve hypotheses.
These links have no obvious or memorable structures and are difficult to track by eye. 
One can add additional visual encoding to these links to depict three types of conclusions (i.e., reject, support, unproven) and conditional dependency.
However, such encoding would further worsen the cluttering of the bipartite graph.
To address this issue, we designed a matrix-based visualisation for HypoML as shown in  \autoref{fig:interface_vertical}(a), where four types of icons (a2) are introduced to indicate reject, support, unproven, and conditional dependency.

%Meanwhile, humans are not good at comparing values (e.g., the accuracy of experiment results, the p-value of statistical comparisons). 
The second shortcoming is that simply listing numerical values (e.g., the accuracy of experiment results, the $p$-value of statistical comparisons) incurs a fair amount of cognitive load upon users who have to compare and analyse them numerically. 
Therefore, we thus visually encoded these values while maintaining the numerical representations. 
In particular, HypoML depicts experiment results with positions, since position is considered to be the most effective visual channel~\cite{munzner2014visualization}.
As shown in ~\autoref{fig:interface_vertical}(c), the position of the circle indicates the average accuracy while the line indicates the 95\% confidence interval.

We decided to encode $p$-value using a glyph, and considered several alternative designs as shown in \autoref{fig:interface_vertical}(b1).
With the first design option, the area of a circle is used to encode the level of statistical significance, i.e., the inverse of a $p$-value.
The less the $p$-value, the more significant the difference, and the larger the circle.
However, in an informal pilot study, this design was found to be ``confusing'' due to the reverse encoding.
With the second design option, the $p$-value is encoded using the area of an orange circle, which is inside a large blue circle of a fixed size.
While this design enables direct observation of statistical significant through the blue area as well as the $p$-value through the orange area, it was found to be ``unintuitive'' for those who were unfamiliar with the definition of $p$-value.
We finally settled down on the third design based on a widely-used illustration for explaining the concept of statistical hypothesis testing.
In this design, the whole shape represents a normal distribution and the area in orange coarsely encodes the $p$-value.
The normal distribution curve can quickly remind users of the meaning of $p$-value.

The third shortcoming is that while depicting the reasoning flow from data to conclusion as in \autoref{fig:workflow} correctly represents the temporal order of the computation, it would slow users down when they wish to find out the conclusions quickly.
We thus reverse the order of the workflow in both the vertical and horizontal versions of the visual user interface (see \autoref{fig:HypoML} and \autoref{fig:interface_vertical}).
The horizontal design is more suitable for wide-screen displays, while the vertical design can be used on portable devices and high-resolution monitors.
Users may benefit from having both designs available.

\begin{figure}[t]
  \centering
  \includegraphics[width=0.9\linewidth]{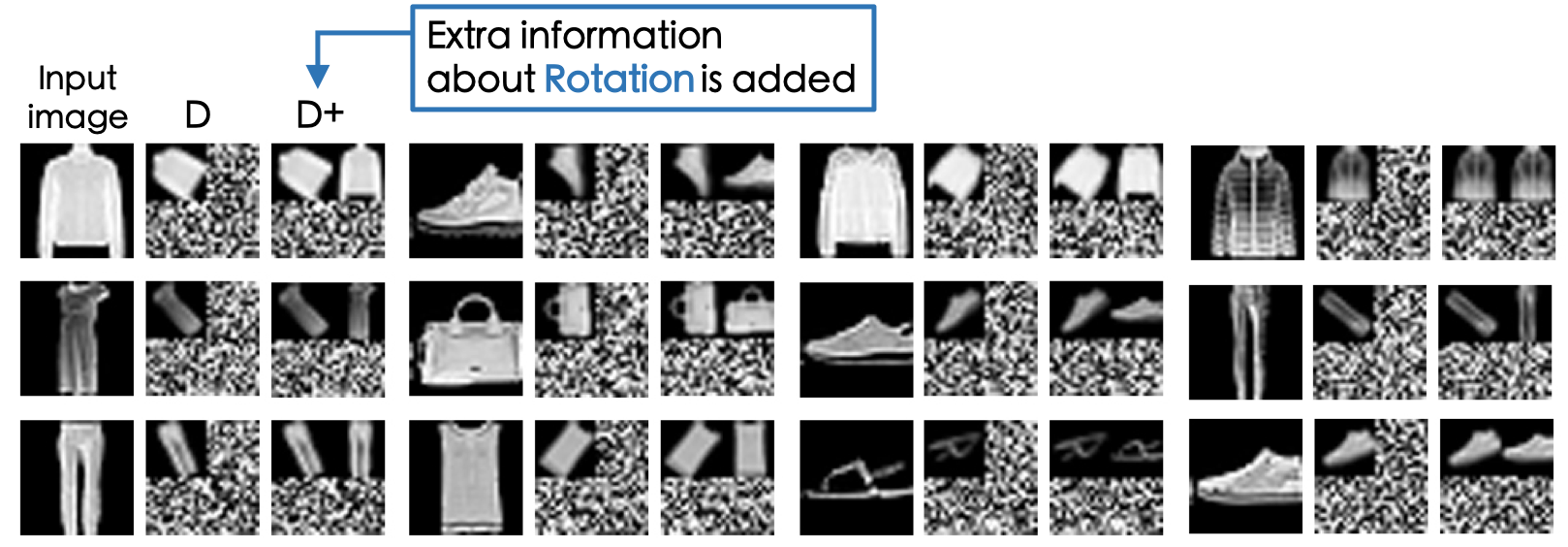}
  \caption{Samples of the training data for testing the concept of rotation correction. For each sample, the left image shows the original object. The middle image shows the corresponding stimulus in the testing dataset $\mathrm{D}$, where the object has been arbitrarily rotated. The right image shows the stimulus in the testing dataset $\mathrm{D+}$ where the rotated object is accompanied by an up-right view of the object.}
  \label{fig:Test1-Data}
  \vspace{-4mm}
\end{figure}

\begin{figure*}[b]
  \centering
  \begin{tabular}{@{}c@{\hspace{4mm}}c@{}}
    \includegraphics[height=40mm]{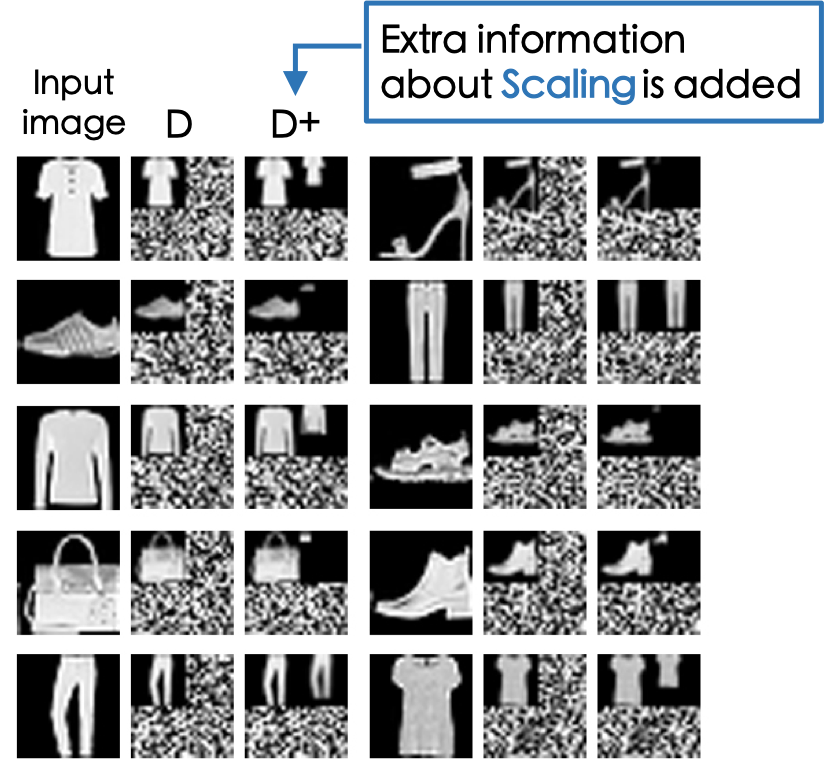}&
    \includegraphics[height=48mm]{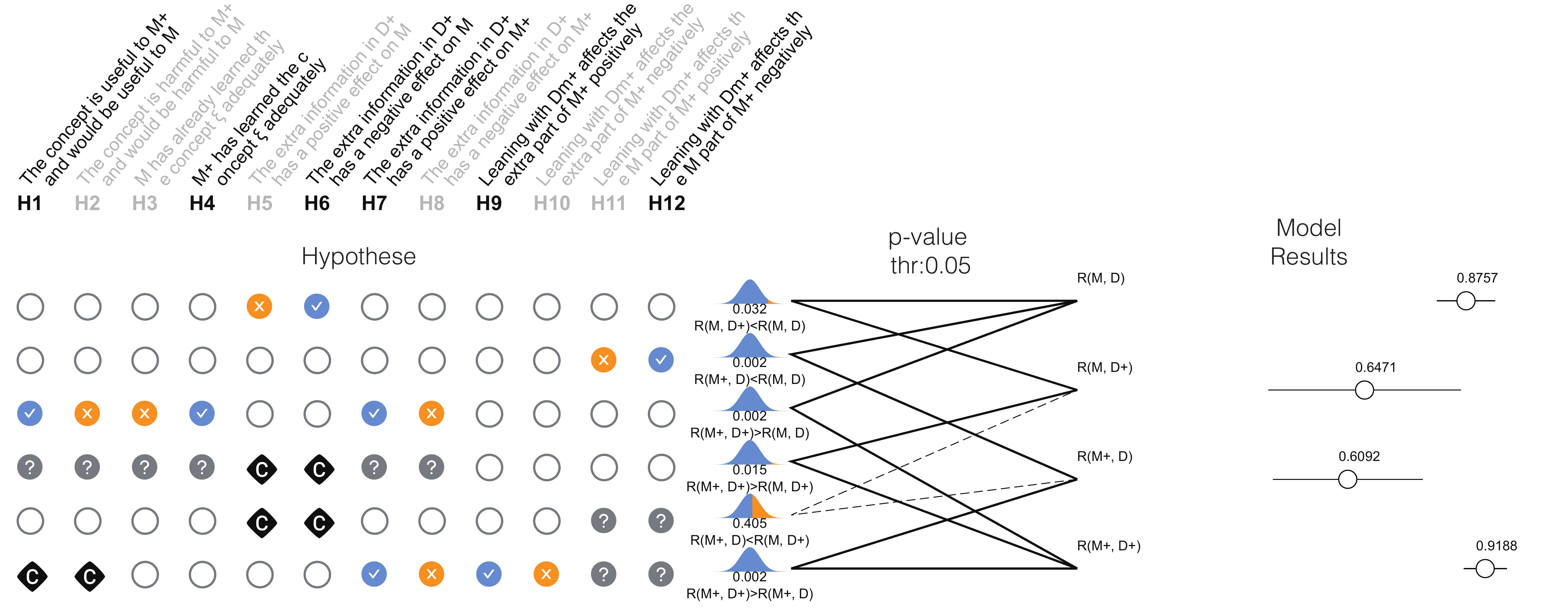}
  \end{tabular}
  \caption{Data samples and the visualization of the testing results for testing the concept of scaling correction. For each sample, the stimulus in $\mathrm{D}$ contains an object of a ``maximized'' size. The stimulus in $\mathrm{D+}$ contains an extra object of a ``relative'' size.}
  \label{fig:Test-S}
  \vspace{-2mm}
\end{figure*}

Both versions of the interface were designed and developed by following an iterative design process with regular feedback from potential users, including model-developers and ML model users.
Through such feedback, we discovered that most users would prefer to observe the conclusions of the hypotheses as soon as the testing results were loaded into HypoML.
They could then decide whether it would be necessary to track back to the statistical comparison and experiment results for detailed reasoning.
We also discovered that double encoding used for the $p$-value and hypotheses had enhanced users' perception of the information and enable them to switch between overview (through visual encoding) and details on demand (through numerical values) rapidly by simply changing their visual attention.
While each $p$-value is already encoded using the glyph and numerical value, we further encode it through its links with the testing results. The link width indicates the reverse of the $p$-value and the link style (i.e., solid or dashed) shows whether the difference between two sets of results is statistical significant or not (\autoref{fig:interface_vertical}(b2)).
While the decision state of a hypothesis is already encoded using icons in the matrix, we double encode it using black and two grey-scale values to the levels of support to the hypothesis (\autoref{fig:interface_vertical}(a1)).
The black color draws users' attention quickly to those hypotheses that have been confirmed.

% To enable effective analysis at different types of screens, we present two versions of the interface, a vertical version (\autoref{fig:interface_vertical}) and a horizontal version (\autoref{fig:HypoML}).
% Which version will be chosen depends on the users' screen.
%Two versions are presented, a vertical version and a horizontal version, to suit different screens.

HypoML supports a set of interactions.
Users are allowed to modify the threshold of $p$-value, which may lead to changes in the conclusions of the hypotheses and dynamical update of the whole visualization.
By hovering on a $p$-value, users can highlight the two corresponding sets of results.

% ====================
% \begin{figure}[t]
%   \centering
%   \includegraphics[width=86mm]{Samples-R-Rotating.jpg}
%   \caption{Samples of the training data for testing the concept of rotation correction. For each sample, the left image shows the original object. The middle image shows the corresponding stimulus in the testing dataset $\mathrm{D}$, where the object has been arbitrarily rotated. The right image shows the stimulus in the testing dataset $\mathrm{D+}$ where the rotated object is accompanied by an up-right view of the object.}
%   \label{fig:Test1-Data}
%   \vspace{-4mm}
% \end{figure}

\section{Results and Discussions}
\label{sec:Results}
% --------------------

% --------------------
% --------------------

\begin{figure*}[t]
  \centering
  \begin{tabular}{@{}c@{\hspace{4mm}}c@{}}
    \includegraphics[height=40mm]{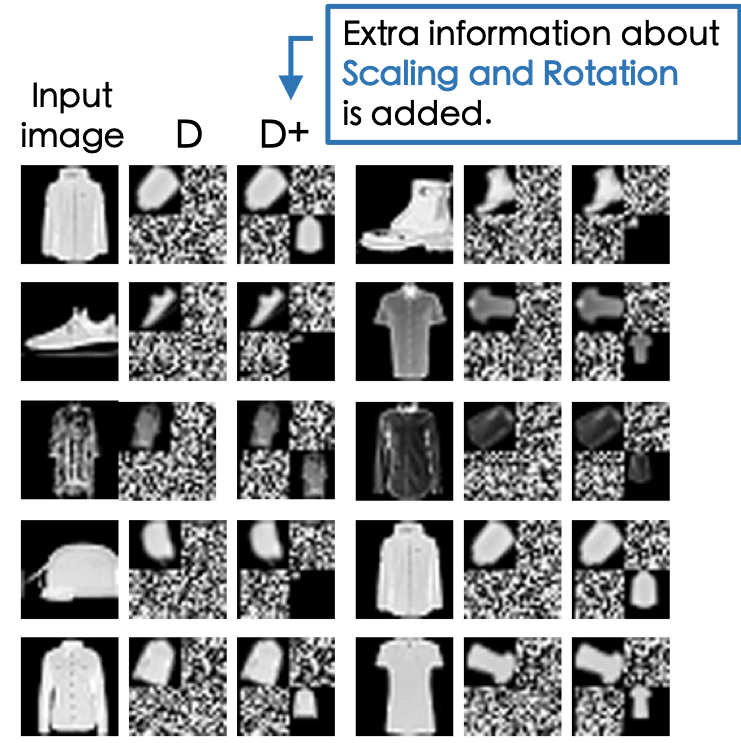}&
    \includegraphics[height=48mm]{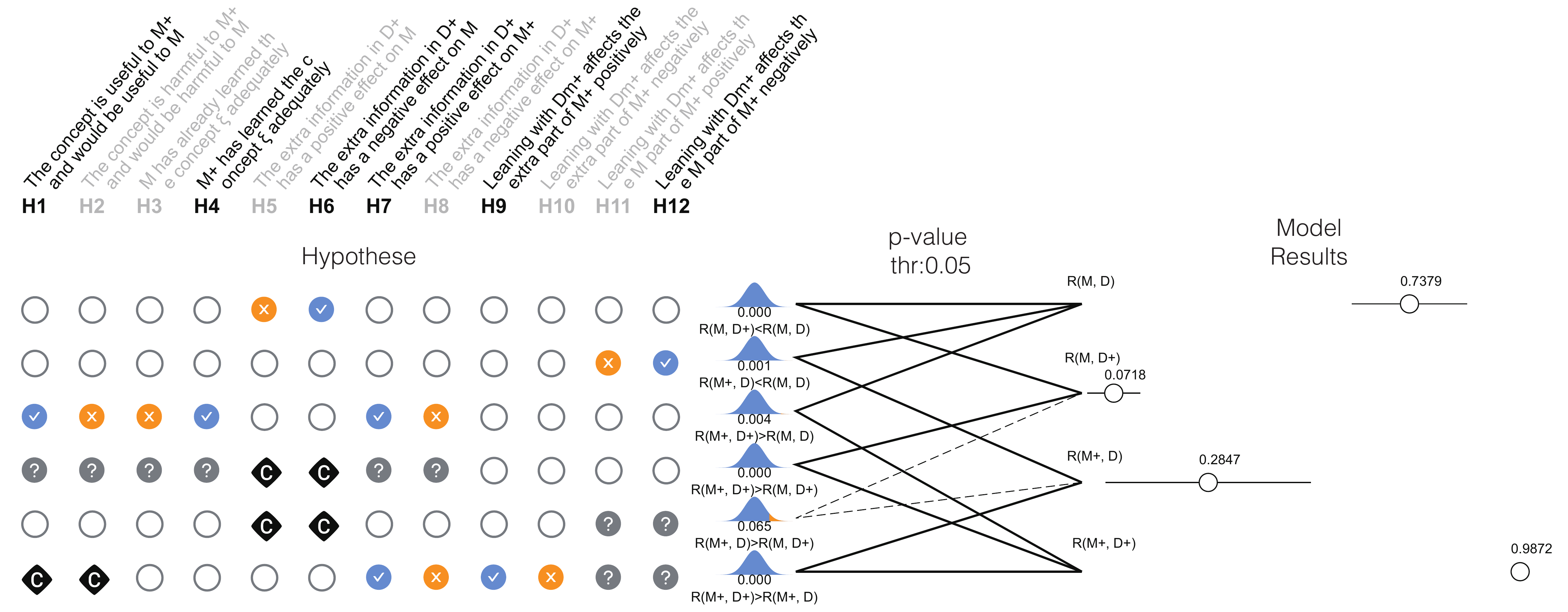}
  \end{tabular}
  \caption{Testing the combined concept of rotating and scaling correction. For each sample, the stimulus in $\mathrm{D}$ contains an object of a rotated and ``maximized'' size. The stimulus in $\mathrm{D+}$ contains an extra object of a ``relative'' size in an up-right view.}
  \label{fig:Test-RS}
\end{figure*}

%----------
% --------------------
\begin{figure*}[t]
  \centering
  \begin{tabular}{@{}c@{\hspace{4mm}}c@{}}
    \includegraphics[height=40mm]{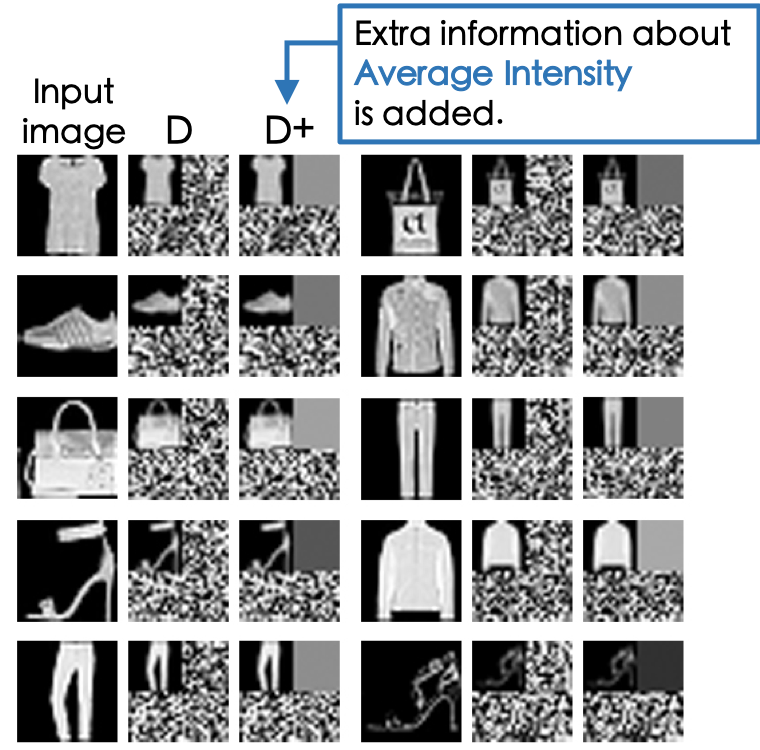}&
    \includegraphics[height=48mm]{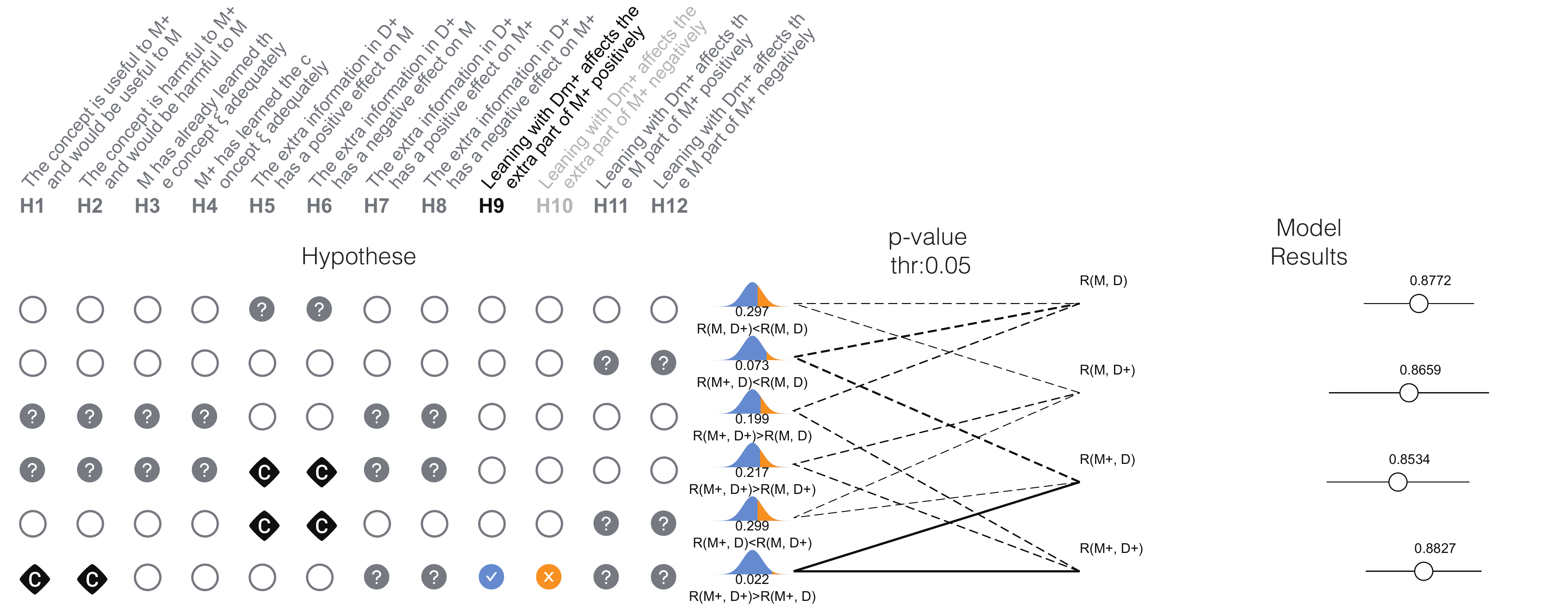}
  \end{tabular}
  \caption{Testing the combined concept of average intensity. For each sample, the stimulus in $\mathrm{D}$ contains an original object. The stimulus in $\mathrm{D+}$ contains an extra piece of information about the average intensity of the object.}
  \label{fig:Test-AvgI}
\end{figure*}

% --------------------

The testing reported in this section is primarily for testing HypoML to see if HypoML can make correct transformation from four sets of results $\mathrm{R_{M,D}}$,  $\mathrm{R_{M,D+}}$, $\mathrm{R_{M+,D}}$, and $\mathrm{R_{M+,D+}}$ to visual representations of the conclusions about 12 hypotheses.
The examples shown are not intended to establish the truth about the goodness of any particular ML technique, but to demonstrate the practical uses of HypoML. 
If a developer suspects an ML model may have a shortcoming, HypoML can help the developer confirm or reject such a hypothesis.
With convolution neural networks (CNN), a common wisdom is that the deeper and the larger a CNN is, more likely a concept will be learned by the CNN.
When our tests show that a particular CNN model has not learned a concept adequately, it does not necessarily mean that a more complicate CNN model would not be able to learn the concept either.
This is indeed what testing is for in software engineering.
The goal of testing is to discover the shortcoming of a model or a piece of software in order to improve the model or software.

We used the Fashion MNIST dataset~\cite{xiao2017fashionmnist} to train a CNN model for classification.
The model was specified using Keras and Tensorflow in Python, and was trained and tested using the Google Colaboratory server.
We use the same CNN structure as that in the official example of Keras.
This CNN consists of the following layers: convolution (3x3x32, RELU), convolution (3x3x64, RELU), max pooling (2x2), dropout (25\%), flatten, dense (128, RULE), dropout(50\%), and dense(10, softmax).
We refer readers to~\cite{keras_example}
for more details.

% It is a simple CNN with two convolution layers, with 32 and 64 nodes, both using relu activation.
% These are followed by MaxPolling, 25\% dropout, flattening, and then a 128 node dense layer.
% There is then another 50\% dropout and a final 10 node/class dense layer with softmax activation.

In each training session, a model is trained using 40,000 training images.
With batch sizes of 128 and 50 epochs, convergence occurs in around 5 minutes.
In each test, a model is tested against 6,666 test images.
These images are all of 28$\times$28 8-bit pixels.
The class labels are: (0) T-shirt/top, (1) Trouser, (2) Pullover, (3) Dress, 
(4) Coat, (5) Sandal, (6) Shirt, (7) Sneaker, (8) Bag, and (9) Ankle boot.

The original images in the Fashion MNIST dataset feature all fashion objects in an upright position.
This naturally leads to a speculation that a trained model may not be rotation invariant.
One possible way to address the need for rotation-invariance is to train a model with images featuring randomly rotated objects, which is widely employed in data augmentation techniques~\cite{simard2003best}.
As humans can determine easily if a fashion object is in an upright position or not, one may hypothesize that a classification model may benefit from the extra information from another model that can detect the rotation angle or perform rotation normalization.

Following the workflow depicted in Figure \ref{fig:Testbed}, we constructed two types of data.
We applied random rotation to each image in the training and testing data.
This resulted in a new training dataset $\mathrm{D}_m$ and testing dataset $\mathrm{D}$.
We then created the $+$ part of the data by simply reusing the original upright images, by presuppose the existence of a rotation normalization model.
% In order to maintain the 28$\times$28 resolution, we downsized the rotated image and the normalized image to 14$\times$14 before configuring the new images.
As illustrated in Figure \ref{fig:Test1-Data}, each group of three images shows an original image (left), an image in $\mathrm{D}_m$ or $\mathrm{D}$ (middle), and an image in $\mathrm{D}_m+$ or $\mathrm{D}+$ (right).
The middle image contains only the rotated image, together with noise in the other three quadrants.
The right image contain both the rotated image and the normalized image, together with noise in the two lower quadrants.

We then trained two models $\mathrm{M}$ and $\mathrm{M}+$, and tested each of them using two datasets $\mathrm{D}$ and $\mathrm{D}+$ according to the workflow in Figure \ref{fig:Testbed}.
From the four sets of testing results, HypoML carries out statistical and logical analysis and displays the results as shown in Figure \ref{fig:HypoML}.
In \autoref{fig:HypoML}, we can oberve that six hypotheses have been confirmed. They indicate:
\begin{itemize}
  \item $\mathbf{H}_1$: The concept of rotation normalization is useful to $\mathrm{M}+$ and would be useful to $\mathrm{M}$.
  \item $\mathbf{H}_3$: $\mathrm{M}+$ has learned from the concept of rotation normalization adequately.
  \item $\mathbf{H}_6$: The extra information in $\mathrm{D}+$, when it is fed to $\mathrm{M}$, has a negative effect on $\mathrm{M}$. Although $\mathrm{M}$ has only learned from noise the upper-right quadrant of the stimuli, when non-noise information appears in that area, it still affects $\mathrm{M}$, in a negative way.
  \item $\mathbf{H}_7$: The extra information in $\mathrm{D}+$ (upper-right quadrant) has a positive effect on $\mathrm{M}+$.
  \item $\mathbf{H}_9$: Learning with $\mathrm{D}_m+$ affects the extra part of $\mathrm{M}+$ positively. This is somehow anticipated because $\mathbf{H}_1$ is confirmed.
  \item $\mathbf{H}_{12}$: Learning with $\mathrm{D}_m+$ affects the $\mathrm{M}$ part of $\mathrm{M}+$ negatively, that is, if the extra information is unavailable, $\mathrm{M}+$ performs worse than $\mathrm{M}$, which has not learned with the extra information. 
\end{itemize}

When working with the dataset, we also noticed that the fashion objects in all images are maximized within the boundary of the image.
We wondered if this would introduce some biases to a trained model.
As humans can usually perceive the size of an everyday object fairly quickly, we hypothesized that a model that can remap a maximized object to a more realistic size may help the classification of such an object.
As shown in Figure \ref{fig:Test-S}, we conducted another test by following the same workflow illustrated in Figure \ref{fig:Testbed}.
In this case, the extra information features a scaled object on the upper-right quadrant.
We measured typical sizes of fashion objects in each category and defined a relative range for the category accordingly.
For the extra information, we randomly selected a scaling factor within the range defined for the corresponding category, and used the factor to scale the image.
The analytical result is shown on the right of Figure \ref{fig:Test-S}.
The conclusions are more or less the same as the hypothesis rotation normalization.

\begin{figure*}[t]
  \centering
  \begin{tabular}{@{}c@{\hspace{4mm}}c@{}}
    \includegraphics[height=40mm]{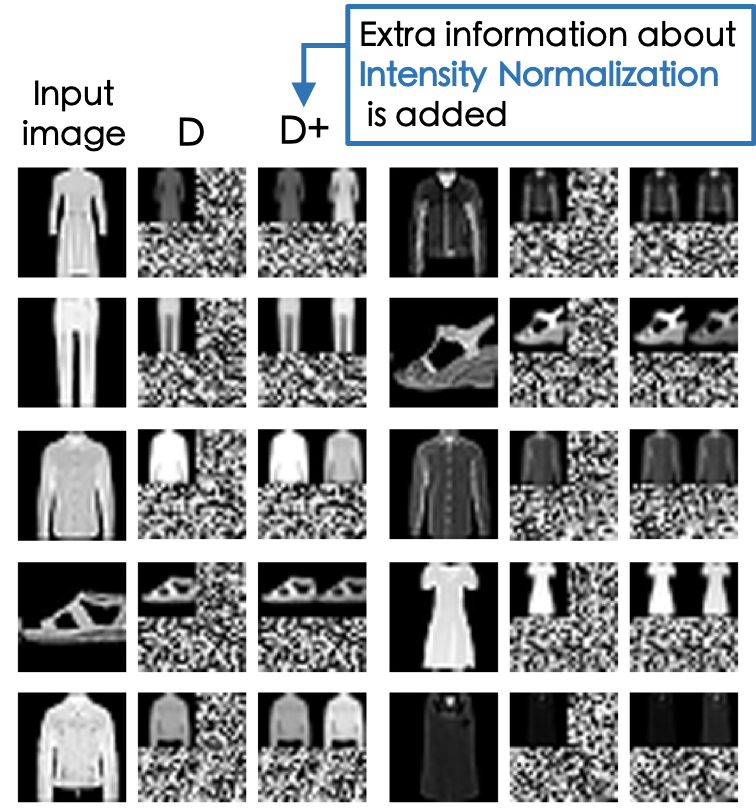}&
    \includegraphics[height=48mm]{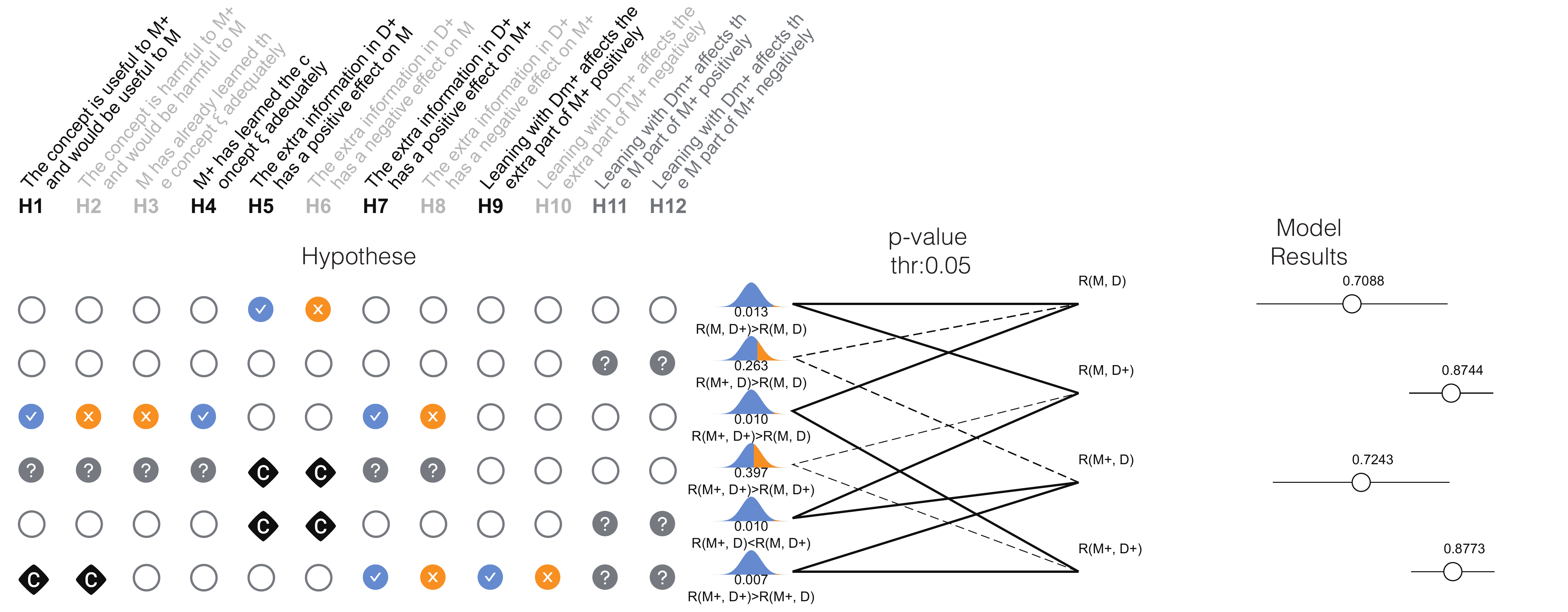}
  \end{tabular}
  \caption{Testing the concept of intensity normalization. For each sample, the stimulus in $\mathrm{D}$ contains an original object whose intensity has been arbitrarily re-scaled. The stimulus in $\mathrm{D+}$ contains an extra object featuring the original intensity as a form of normalization.}
  \label{fig:Test-NormI}
\end{figure*}

To demonstrate a slightly more complex design of a test, we combined the above two tests to examine the combined effects of the two concepts, namely rotation normalization and relative scaling.
As shown in Figure \ref{fig:Test-RS}, we used the upper-left quadrant for the rotated object as the information present in all training and testing data.
We placed rotation-normalized and relatively-scaled object at the lower-right quadrant.
As perhaps expected, the test confirmed the same set of hypotheses as the two tests mentioned before.

In general, a CNN is expected to learn features about some aggregated properties (e.g., mean, median, or mode). We thus conducted a test to see whether providing such a feature as a piece of extra information is useful.
As shown in Figure \ref{fig:Test-AvgI}, we introduced the average intensity value of an object as a single-colored square in the upper-right quadrant.
The analysis of the test results indicates that most hypotheses are unproven. In other words, we cannot be sure if this extra piece of information is useful or harmful.
The only hypothesis that has been confirmed is $\mathbf{H}_9$, i.e., learning with $\mathrm{D}_m+$ affects the extra part of $\mathrm{M}+$ positively.
However, this does not translate to a confirmation of $\mathbf{H}_7$ about the overall positive impact to $\mathrm{M}+$.
By observing the details about how this hypothesis (i.e.,  $\mathbf{H}_9$) was confirmed, we can see that it is confirmed only within the context of $\mathrm{M}+$, without involving any tests about $\mathrm{M}$.

Considering further about the intensity of the images, one common idealized requirement in computer vision is lighting invariance, i.e., a model can recognize the same object under different lighting conditions.
We thus hypothesized that another model for normalizing the intensity of an image may help a classification model.
Using a similar strategy as in the first test (random rotation), we randomly change the intensity of the original images to create the benchmark datasets $\mathrm{D}_m$ and $\mathrm{D}$.
We then use the original images as the extra information, presupposing that the original images were the results of intensity normalization.

Figure \ref{fig:Test-NormI} shows that the extra information is useful to $\mathrm{M}+$ ($\mathbf{H}_1$), and $\mathrm{M}+$ has learned the concept adequately ($\mathbf{H}_4$).
While the test confirms $\mathbf{H}_7$ and $\mathbf{H}_9$, it is inconclusive about $\mathbf{H}_{11}$ and $\mathbf{H}_{12}$.
Interestingly, the test confirms $\mathbf{H}_5$ unexpectedly, i.e., the extra information in $\mathrm{D}+$ has a positive effect on $\mathrm{M}$.
This is in some way related to the failure to confirm $\mathbf{H}_{12}$ as in some earlier tests.
For each image in $\mathrm{D}+$, the signals in the extra information (i.e., the upper-right quadrant), which in many ways is similar to those in $\mathrm{D}$ (i.e., upper-left quadrant).
One possible explanation is the signals in the upper-right quadrant somehow strengthen the signals in the upper-left quadrant, even though $\mathrm{M}$ has not learned to use the extra information.

We have also conducted several other tests about the randomly-sized class labels and images with incorrect labels.
HypoML has also shown to be useful for support such hypothesis testing.

% ====================
\section{Conclusions}
\label{sec:Conclusions}
In this paper, we propose a novel testing framework to aid the evaluation of ML models. 
% In particular, this framework tests a set of hypotheses about if an ML model can benefit from extra information about a concept, and if so, how the extra information may affect the models trained with and without such extra information.
In particular, this framework tests a set of hypotheses about a concept, checking whether extra information about the concept can benefit an ML model, and if so, how the extra information affects the model.
The testing framework is underpinned by statistical analysis of the experiment results as well as logical inferences about the relations between six statistical conclusions and twelve hypotheses.
Through an implementation of this framework HypoML, we demonstrate that with a purposely-designed visual representation, model-developers can visualize the conclusions about the twelve hypotheses as soon as the four sets of testing result data become available.
This approach complements the traditional way of observing various plots for monitoring neuron activities, such as activation plots and gradient ascent plots.
Model-developers, who observe any interesting patterns or failed to find desired patterns, can now formulate a concept-based hypothesis and carry out a structured test to evaluate their hypotheses.

We recognize that HypoML is only one of the many steps towards an ultimate goal of developing a powerful testing suite for evaluating, understanding, and explaining ML models.
There is a need for further theoretical and practical developments in this direction, including, for instance, formulating more detailed logical analysis for sub-group analysis of the testing results, designing an advanced user interface for supporting detailed observation of sub-group analysis, and integrating with other visualization techniques for observing, understanding, and explaining ML models.    

%-------------------------------------------------------------------------

% bibtex
\bibliographystyle{eg-alpha-doi}
\bibliography{HypoML}
% biblatex with biber
% \printbibliography                

\end{document}